\documentstyle[12pt]{article}

\begin{document}
\newcommand {\be}{\begin{equation}}
\newcommand {\ee}{\end{equation}}
\newcommand {\bea}{\begin{array}}
\newcommand {\cl}{\centerline}
\newcommand {\eea}{\end{array}}
\baselineskip 0.65 cm
\begin{center}
\bf \Large {Different D-brane Interactions}
\end{center}
\cl {\it H. Arfaei}
\cl {\it and}
\cl{\it M.M. Sheikh Jabbari}
\cl {\it Institute for Studies in Theoretical Physics and Mathematics IPM}
\cl{\it P.O.Box 19395-5746, Tehran,Iran}
\cl{\it Department of Physics Sharif University of Technology}
\cl{\it P.O.Box 11365-9161}
\cl{\it e-mail  Arfaei@ theory.ipm.ac.ir}
\cl{\it e-mail  jabbari@netware2.ipm.ac.ir}
\vskip 2cm
\begin{abstract}
We use rotation of one D-brane with respect to the other to reveal the hidden 
structure of D-branes in type-II theories. This is done by calculation of the 
interaction amplitude for two different parallel and angled branes. 
The analysis of strings with different boundary conditions at the ends 
is also given. 
The stable configuration for two similar branes occurs when they are anti-parallel. 
For branes of different dimensions stability is attained for either parallel or 
anti-parallel configurations and when dimensions differ by four the 
amplitude vanishes at the stable point. The results serve as more evidence that
D-branes are stringy descriptions of non-perturbative extended solutions of
SUGRA theories, as low energy approximation of superstrings.
\end{abstract}
\newpage
{\it I. Introduction}

 The observation that D-branes provide stringy descriptions of topological
objects in string theory has been source of many recent developments [1,2,3]. They
are soureces of $RR$ charges and are argued to provide short distance probes in
some special cases, where strings fail [4,5]. Like solitonic solutions they
interact via exchange of the basic quanta of the theory i.e. closed strings which
are intrepreted as D-strings in the dual channel.

It is important to verify that the D-branes which are topological defects, able to trap the end of the strings, and solitonic black solutions of the 
low energy limit which are non-perturbative objects of the theory are identical 
objects. If this is so we have our hands on a stringy description of a
non-perturbative effects without having a non-perturbative formalism.

There have been attempts to present justifications of this view [6]. The main 
contribution coming from work of Polchinski [1,2], where  it has been 
shown that D-branes carry $RR$ charges. It was shown that the gravitional interaction
of two D-branes is balanced by their $RR$ repulsion. On the other hand
solitonic or black branes solutions to SUGRA [7,8] have more structure than D-branes which are merely 
given by hyperplanes.

In this work we consider the interaction of two different D-branes which are not
parallel and find the orientation dependence of their interaction. In Nilsen-Olsen vortices
( or similarly in typeII super conductivity), rotating a vortex by 180 degrees, the 
repulsive force of the parallel flux lines turns into attractive force of the
two anti-parallel vortices. We see similar effects for the $RR$ force between
two non-parallel D-branes and find its orientation dependence [9] and furthermore 
the angle between D-branes can be considered as a collective dynamical varible.
The stable situation is where two branes are anti-parallel,
which is justified by field theoretical arguments. This observation gives further
evidence for the identity of D-branes and solitonic black branes, where their
interactions is described by $\ \int J.A \ $ term in which  $J$ 
is the related P-brane current. We also find the velocity dependence of the 
interactions of two D-branes which confirms the intuitive expectation to see 
the corresponding magnetic $RR$ force in agreement with the low velocity results of [4] and [6].  
For completeness we also include the ordinary bosonic strings results in the  
appendix, in this case only graviton and dilaton are present. 

{\it II. Interaction of two parallel D-branes of P and $P'$ dimensions in type-II theories:}

Consider two D-branes of dimensions $p$ and $p'$ ($p \geq p'$).
We assume that they interact via exchange of closed strings, i.e. the basic object
of the theory. Following [1] to calculate the amplitude we consider the open
string loop in the dual channel. In this part we assume the branes to be parallel
and non intersecting. The case of angled (non-parallel non-intersecting) branes 
will be considered later. The $D_p$-brane is located at $X^{\mu}=Y^{\mu} \ \ \ \mu=p+1,...,9 \ \ $
and the $D_{p'}$-brane at $X^{\mu}=0 \ \ \ \mu=p'+1,...,9  \ $.

We realize three types of open D-strings [10,11]: $i$)both ends on $D_{p}$-brane, $ii$)
both ends on $D_{p'}$-brane and $iii$) one end on $D_{p}$ and the other on 
$D_{p'}$-brane. The first two types represent the excitations of the individual 
branes but the third is responsible  for their interaction. The first $p'$ components
satisfy Neumann boundary conditions at both ends, the $\Delta=p-p'$ components  
satisfy Neumann at $\sigma=0$ and Dirichlet at $\sigma=\pi$ and remaining $9-p$ 
components Dirichlet at both ends. This leads to half integer modes for the 
$X^{\mu} \ \ \mu=p'+1,...,p\ $ which are not present in $p$ and $p$ case. 
Hence the open strings stretching between two branes satisfy the conditions:
\be
\sigma=0 \left\{  \begin{array}{cc}
X^{\mu}=0 & \mu=p+1,..,9 \\ 
\partial_{\sigma}X^{\mu}=0  &  \mu=0,..,p
\end{array}\right.
\ee
\be
\sigma=\pi \left\{  \begin{array}{cc}
X^{\mu}=Y^{\mu} & \mu=p'+1,..,9 \\ 
\partial_{\sigma}X^{\mu}=0  &  \mu=0,..,p'
\end{array}\right.
\ee
The boundary conditions for the world-sheet fermions $\psi_+^{\mu}$and $\psi_-^{\mu}$
follows from world-sheet super symmetry transformation:
\be
\delta X^{\mu}=\bar \epsilon \psi^{\mu}
\ee
where $ \bar \epsilon$ is a 2-D world-sheet fermion.
There are, as expected, two types of open strings, Ramond($R$) type and Neveu-Schwarz($NS$)type. 
Putting all these together mode expansion for $X^{\mu}$ are:
\be
\begin{array}{cc}
 X^{\mu}=p^{\mu}\tau +\sum_{n \in Z} {1\over n} \alpha_n^{\mu} e^{-in\tau} \cos n\sigma &  \mu= 0,...,p' \\
 \ \ \ \ \ =\sum_{r \in Z+1/2} {1\over r} \alpha_r^{\mu} e^{-ir\tau} \sin r\sigma &  \mu=p'+1,...,p  \\      
 \ \ \ \ =Y^{\mu} {\sigma\over \pi}+\sum_{n \in Z} {1 \over n} \alpha_{n}^{\mu}e^{-in\tau} \sin n\sigma &  \mu=p+1,...,9. 

\end{array}
\ee
for both $R$ and $NS$ types, and for the $\psi_{\pm}^{\mu}$:                           
\be
\left\{ \bea{cc}
\psi^{\mu}_+=\sum_{n \in Z} d^{\mu}_n e^{-in(\tau+\sigma)} \;\;\; 
\psi^{\mu}_-=\sum_{n \in Z} d^{\mu}_n e^{-in(\tau-\sigma)} \;\;\;\mu=0,...,p'  \\
\psi^{\mu}_+=\sum_{n \in Z} d^{\mu}_n e^{-in(\tau+\sigma)} \;\;\; 
\psi^{\mu}_-=-\sum_{n \in Z} d^{\mu}_n e^{-in(\tau-\sigma)} \;\;\;\mu=p+1,...,9 \\
\psi^{\mu}_+=\sum_{r \in Z+1/2} d^{\mu}_r e^{-ir(\tau+\sigma)} \;\;\; 
\psi^{\mu}_-=-\sum_{r \in Z+1/2} d^{\mu}_r e^{-ir(\tau-\sigma)} \;\;\;\mu=p'+1,...,p 
\eea \right.
\ee
for the {\bf{R}}-sector and
\be
\left\{ \bea{cc}
\psi^{\mu}_+=\sum_{r \in Z+1/2} b^{\mu}_r e^{-ir(\tau+\sigma)} \;\;\; 
\psi^{\mu}_-=\sum_{r \in Z+1/2} b^{\mu}_n e^{-ir(\tau-\sigma)} \;\;\;\mu=0,...,p'  \\
\psi^{\mu}_+=\sum_{r \in Z+1/2} b^{\mu}_r e^{-ir(\tau+\sigma)} \;\;\; 
\psi^{\mu}_-=-\sum_{r \in Z+1/2} b^{\mu}_n e^{-ir(\tau-\sigma)} \;\;\;\mu=p+1,...,9 \\
\psi^{\mu}_+=\sum_{n \in Z} b^{\mu}_n e^{-in(\tau+\sigma)} \;\;\; 
\psi^{\mu}_-=-\sum_{n \in Z} b^{\mu}_n e^{-in(\tau-\sigma)} \;\;\;\mu=p'+1,...,p 
\eea \right.
\ee
for the {\bf{NS}}-sector.
Upon quantization we get the following commutation relations:
\be 
[\alpha_r^{\mu},\alpha_s^{\nu}]=\delta_{r+s} \delta^{\mu \nu}
\ee
\be 
\{d_r^{\mu},d_s^{\nu}\}=\delta_{r+s} \delta^{\mu \nu}
\ee
\be 
\{b_n^{\mu},b_m^{\nu}\}=\delta_{n+m} \delta^{\mu \nu}
\ee
Commutations of the other components are the same as usual. The mass spectrum
of the strings are given by:
\be
\alpha'M^2=\frac{Y^2}{4\pi^2\alpha'}+N-({1 \over 2}-\frac{\Delta}{8}) \;\;\;\;\, 
\Delta=p-p'
\ee
\be
N=\sum_{n>0}^{8-\Delta} \alpha_{-n}.\alpha_{n}+\sum_{r>0}^{\Delta} \alpha_{-r}.\alpha_{r}+
\sum_{r>0}^{8-\Delta} rb_{-r}.b_{r} + \sum_{n>0}^{\Delta} nb_{-n}.b_{n}. 
\ee
for the $NS$ sector and by
\be
\alpha'M^2=\frac{Y^2}{4\pi^2\alpha'}+N \;\;\;\;\,\Delta=p-p'
\ee
\be
N=\sum_{n>0}^{8-\Delta} \alpha_{-n}.\alpha_{n}+\sum_{r>0}^{\Delta} \alpha_{-r}.\alpha_{r}+
\sum_{n>0}^{8-\Delta} nd_{-n}.d_{n} + \sum_{r>0}^{\Delta} rd_{-r}.d_{r}. 
\ee
for $R$ sector, $\sum^{8-\Delta} $ and $\sum^{\Delta}$ mean that the 
sums include $(8-\Delta)$ and $\Delta$ component, which are related
to $DD$ and $NN$ components and $DN$ components respectively.

The amplitude for the exchange of one closed string expressed in terms of the open  
string loop in the dual channel is:
\be 
A=\int {dt \over 2t} \sum_{i,p}e^{-2\pi\alpha' t (p^2+M_i^2)}
\ee 
where $i$ indicates the modes of the open string and $p$ the momentum which has non-zero
components in the first $p'+1$ dimensions.

Before calculating the amplitude we must ensure that in closed string channel no
tachyon will emerge. The following generalized GSO projections [12] are consistently sufficient 
for this purpose:
\be 
NS:\ \ \ P=1/2(1-G) ;\ \ \ G=(-1)^{\sum_{r>0}^{8-\Delta} b_{-r}b_{r} + \sum_{n\geq0}^{\Delta} b_{-n}b_{n}}.  
\ee
\be
R:\ \ \ P=1/2(1+\Gamma)  ;\ \ \ \Gamma=(-1)^{\sum_{n\geq 0}^{8-\Delta} d_{-n}d_{n} + \sum_{r>0}^{\Delta} d_{-r}d_{r}}.
\ee
The GSO again suggests that, the $NS$  contribution 
must be subtracted from the  $R$ contribution, thus we obtain:
\be
 A=2V_{p'+1} \int \frac{dt}{2t}(8\pi^2\alpha't)^{-(p'+1)/2}e^{-\frac{Y^2t}{2\pi^2\alpha'}} 
 (\bf{ NS-R}). 
\ee
where $\bf{ NS}$ and $\bf{ R}$ are given by 
\be
{\bf NS} = 2^{\Delta/2-1}\ q^{-1+{\Delta \over 4}}
\biggl(\prod{\frac{(1-q^{2n})}{(1-q^{2n-1})}}\biggr) ^{\Delta}
\biggl(\prod{\frac{(1+q^{2n})}{(1+q^{2n-1})}}\biggr) ^{\Delta}
\biggl(\prod{\frac{(1-q^{2n})}{(1+q^{2n-1})}}\biggr) ^{-8}
\ee
\be
{\bf R}= 2^{3-\Delta/2}
\biggl(\prod{\frac{(1-q^{2n})}{(1-q^{2n-1})}}\biggr) ^{\Delta}
\biggl(\prod{\frac{(1+q^{2n-1})}{(1+q^{2n})}}\biggr) ^{\Delta}
\biggl(\prod{\frac{(1-q^{2n})}{(1+q^{2n})}}\biggr)   ^{-8}
\ee

When examined in the closed string channel we see that no $RR$ particle is
exchanged due to the fact that the term coming from  $(-)^G$ is absent here.
The interaction comes from the exchange of $NSNS$ particle where in the small
$t$ limit consists of graviton and dilaton [6]. 

Lack of $RR$ term agrees with our field theoretic intuition. 
If $D_p$-brane is the manifestation of solitonic black branes their $RR$ field
is described by a $p+1$ from. This potential can be probed by a $RR$ charge of 
the same type, i.e. only by a P-brane and not $P'$-branes.

If we take the small $t$ limit to isolate the contribution of the massless closed
strings, long range interaction amplitude is obtained to be:

\be
A= V_{p'+1}(4\pi^2\alpha')^{3-{p+p' \over 2}}\ {(2-\Delta/2)}\pi 
G_{9-p}(Y^2),
\ee
in agreement with result of [2] that $T_p=(4\pi^2\alpha')^{3-p}$.

Note that when $\Delta=4$ the amplitude $vanishes\ in\ all\ orders\ of\ t $, 
indicating the existence of SUSY [12] and $stable$ classical BPS solutions with non-vanishing two 
$different$ $RR$ charges.

{\it III.Angled branes in type-II theories:}

 In this part we consider two non-intersecting D-branes which make an angle in the
plane of (p,p+1). Thus the second brane is like the first one but it is rotated by the angle $\pi\theta$
in the (p,p+1) plane. The boundary conditions for the $X^{\mu}$ are:
\be
\sigma=0 \left\{  \begin{array}{cc}
X^{\mu}=0 & \mu=p+1,..,9 \\ 
\partial_{\sigma}X^{\mu}=0  &  \mu=0,..,p
\end{array}\right.
\ee
\be
\sigma=\pi \left\{  \begin{array}{cc}
X^{\mu}=Y^{\mu} & \mu=p+2,..,9 \\ 
\partial_{\sigma}X^{\mu}=0  &  \mu=0,..,p-1 \\
X^{p} \sin{\pi \theta}-X^{(p+1)}\cos{\pi \theta} =0 \\
\partial_{\sigma}X^{p}\cos {\pi \theta}+ \partial_{\sigma}X^{(p+1)}\sin {\pi \theta}=0
\end{array}\right.
\ee
where $\pi\theta$ is the angle between two non-intersecting $0 \leq \theta \leq 1$ $D$-branes.
The world-sheet fermion boundary conditions again are obtained using the world-sheet
SUSY transformation written above and the mode expansions are the same as usual except
in the (p,p+1) components where the modes look like twisted string modes with indices ranging 
over $Z\pm \theta$: 
\be
\begin{array}{cc}
 \ \ \ \ X^p =\sum {1 \over n_+} \alpha_{n_+}^p e^{-in_+\tau} \cos n_+ \sigma +
 \sum {1 \over n_-} \alpha_{n_-}^p e^{-in_-\tau} \cos n_- \sigma. \\
\ \ \ \ X^{(p+1)}= \sum {1 \over n_+} \alpha_{n_+}^p e^{-in_+\tau} \sin n_+ \sigma -
\sum {1 \over n_-} \alpha_{n_-}^p e^{-in_-\tau} \sin n_- \sigma. 
\end{array}
\ee
\be
\left\{ \bea{cc}
\psi^{p}_+=\sum_{n_+}  d^{p}_{n_+}e^{-in_+(\tau+\sigma)} \;\;\; 
\psi^{p}_-=\sum_{n_+}  d^{p}_{n_+}e^{-in_+(\tau-\sigma)} \;\;\; 
\\
\psi^{p+1}_+=i\sum_{n_+}  d^{p}_{n_+}e^{-in_+(\tau+\sigma)} \;\;\; 
\psi^{p+1}_-=-i\sum_{n_+}  d^{p}_{n_+}e^{-in_+(\tau-\sigma)} \;\;\; 
\eea\right.
\ee
\be
\left\{ \bea{cc}
\psi^{p}_+=\sum_{n_-}  d^{p}_{n_-}e^{-in_-(\tau+\sigma)} \;\;\; 
\psi^{p}_-=\sum_{n_-}  d^{p}_{n_-}e^{-in_-(\tau-\sigma)} \;\;\; 
\\
\psi^{p+1}_+=-i\sum_{n_-}  d^{p}_{n_-}e^{-in_-(\tau+\sigma)} \;\;\; 
\psi^{p+1}_-=i\sum_{n_-}  d^{p}_{n_-}e^{-in_-(\tau-\sigma)} \;\;\; 
\eea\right.
\ee
for the $R$ sector, where $n\pm$ stands for $n\pm \theta$. 
$NS$ sector can be obtained by changing $n_\pm$ to $r_\pm$,
where r is in $Z+1/2$.
 The commutation relation of the twisted operators reads as:
\be
\bigl[\alpha_{-n\pm},\alpha_{-n'\mp}\bigr]=n_{\pm}\delta_{n+n'}
\ee

If we insert these commutations the corresponding mass spectrum becomes as:
\be
\alpha'M^2=\frac{Y^2}{4\pi^2\alpha'}+N(\theta)
\ee
$$
N(\theta)=\sum_{n>0}\alpha_{-n}.\alpha_{n}+
\sum_{n>0} nd_{-n}d_n + \sum_{n>0}\alpha_{-n_+}^p.\alpha_{n_+}^p+ \sum_{n>0}\alpha_{-n_-}^p.\alpha_{n_-}^p
$$
\be
+\sum_{n>0}n_+ d_{-n_+}^p.d_{n_+}^p+ \sum_{n>0}n_- d_{-n_-}^p.d_{n_-}^p
+{1 \over 2}( \alpha_{\theta}\alpha_{-\theta}+ \alpha_{-\theta}\alpha_{\theta} )
+{1 \over 2}\theta( d_{-\theta}^p d_{\theta}^p- d_{\theta}^p d_{-\theta}^p )
\ee

for the $R$ sector , d and $\alpha$ are similar to what was defined before.

In \ $NS$ \ sector the oscillator part is the same as \ $R$ \ sector but \ $d$ \ is 
replaced by \ $b$ \ , \ $n$ \ by \ $r$ \ and but the ordering constant is $1/2$,
as we see the ordering constant doesn't depend on $\theta$ in both \ $R$ \ ,\ $NS \ $
cases.
The GSO projection is again needed to remove tachyons from the closed string channel.
This is implemented by using:
\be 
G=(-1)^{\sum_{r> 0} b_{-r}b_{r} + \sum_{r\geq 0} b_{-r_+}^p.b_{r_+}^p+ \sum_{r\geq 0} b_{-r_-}^p.b_{r_-}^p}.
\ee
\be 
\Gamma=(-1)^{\sum_{n\geq 0} d_{-n}d_{n} + \sum_{n\geq 0} d_{-n_+}^p.d_{n_+}^p+ \sum_{n\geq 0} d_{-n_-}^p.d_{n_-}^p}
\ee
So amplitude is:
\be
 A(\theta)=2V_{p} \int \frac{dt}{2t}(8\pi^2\alpha't)^{-p/2}e^{-\frac{Y^2t}{2\pi^2\alpha'}} 
 (\bf{ NS-R}). 
\ee
where $\bf{ NS,R}$ are given by
\be
{\bf R}=8\pi^3 \biggl(\frac{\Theta_2(0\mid it)}{\Theta'_1(0\mid it)}\biggr)^3 \;\ 
\frac{i\Theta_2(it\theta \mid it)}{\Theta_1(it\theta \mid it)}
\ee
\be
{\bf NS}=8\pi^3 \biggl[\biggl(\frac{\Theta_3(0\mid it)}{\Theta'_1(0\mid it)}\biggr)^3 \ 
\frac{i\Theta_3(it\theta \mid it)}{\Theta_1(it\theta \mid it)}-
\biggl(\frac{\Theta_4(0\mid it)}{\Theta'_1(0\mid it)}\biggr)^3 \ 
\frac{i\Theta_4(it\theta \mid it)}{\Theta_1(it\theta \mid it)}\biggr]
\ee
To analyze this  result consider the massless contribution by looking the
small $t$ limit:
$$
 A(\theta)=V_{p} \int \frac{dt}{t}(8\pi^2\alpha't)^{-p/2}e^{-\frac{Y^2t}{2\pi^2\alpha'}} 
 \bigl(8t^3 \tan(\pi\theta/2)\sin^2(\pi\theta/2)\bigr)
$$ 
\be 
 \ =4V_p\tan(\pi\theta/2)\sin^2(\pi\theta/2)(4\pi^2\alpha')^{3-p} \ G_{8-p}(Y^2).
\ee
\newpage
Following comments are in order:
  
  1. At $\theta=0$ $\ A\ $ vanishes( reproducing the result of [1]).
  
  2. The potential altogether is proportional to $(1-\cos\pi\theta)^2$, 
which has a minimum at $\theta=1$ and a maximum at $\theta=0$.
This shows that, two P-branes tend to rotate so that they become anti-parallel.
The anti-parallel case was also considered in [9], whose result is a special case of ours.
It is instructive to compare the type-II results with the bosonic one which are 
given in the appendix. In type-II the $RR$ interaction of the D-branes reveal their
structure more which is not obvious in their simple definition as hyperplanes.
In the bosonic thoery the amplitude is symmetric under $\theta \leftrightarrow 1-\theta$
since the structure of the extended object is invariant under parity.

  3. $RR$ and $NSNS$ massless contributions are:
\be
V_{RR}=8V_0 \cos\pi\theta  \;\;\;\;\ V_0= 4V_p{1 \over  \sin\pi\theta }(4\pi^2\alpha')^{3-p} \ G_{8-p}(Y^2).
\ee
\be
V_{NSNS}=-4V_0 (1+\cos^2\pi\theta )            
\ee
\be
V_{tot}=V_{NSNS}+ V_{RR}=-V_0(1-\cos\pi\theta )^2.     
\ee
 The $\theta$ dependence of the amplitude may be considered as further evidence 
for the identity of D-brane and extended low energy topological objects. The  
black branes have non-zero $RR$ gauge fields  which are  proportional 
to  $\epsilon^{\mu_0 ...\mu_p}\ $ 
in the $p+1$ dimensional P-brane hyperplane. The charge density is 
proportional to  
 $\epsilon $ in the orthogonal space hence having an $orientation$.
Close examination of the $\int J.A $ term as the interaction, clarifies the  
$\cos \pi\theta \ $ dependence of the $V_{RR}$.

Although the stringy description of D-branes is very simple with no input, 
the structure has become clear under rotation of one of them.

 4. If we look at $\theta$ as a collective coordinate the amplitude suggests that 
two D-branes system will oscillate around the equilibrium position with frequency proportional to the $RR$ charge 
density(equal to the D-brane tension).

If $\theta\ $ is fixed the oscillator modes of $X$ corresponding to $\alpha_{\theta}$
and $\alpha_{-\theta}$ shows excitations with frequency proportional to $\theta$ since:
\be
\bigl[\alpha_{\theta},\alpha_{-\theta}\bigr]=\theta
\ee
These oscillations are much more easily excited than the collective mode oscillations
of $\theta$,which requires infinite energy if the space is not compactified along 
$(p,p+1)$ plane. The limit $\theta$ going to zero is also interesting, in this limit
only relevent part is $n=0$ term in summitions
\be
\begin{array}{cc}
X^p ={1 \over 2\theta} \alpha_{\theta}^p e^{-i\theta\tau} \cos \theta\sigma -
{1 \over 2\theta } \alpha_{-\theta}^p e^{i\theta\tau} \cos \theta\sigma+ oscil. \\
\ \simeq {1 \over 2\theta}\bigl(\alpha_{\theta}^p-\alpha_{-\theta}^p\bigr)-   
i\bigl(\alpha_{\theta}^p+\alpha_{-\theta}^p\bigr)\tau+oscil.
\end{array}
\ee
\be
\bea{cc}
X^{(p+1)}= {1 \over 2\theta} \alpha_{\theta}^p e^{-i\theta\tau} \sin \theta\sigma -
{1 \over 2\theta } \alpha_{-\theta}^p e^{i\theta\tau} \sin \theta\sigma+ oscil. \\
\simeq \bigl(\alpha_{\theta}^p-\alpha_{-\theta}^p\bigr)\sigma+oscil.
\eea
\ee
and as it's seen from commutation relations:
\be
[{1 \over \theta}\bigl(\alpha_{\theta}^p-\alpha_{-\theta}^p\bigr),  
i\bigl(\alpha_{\theta}^p+\alpha_{-\theta}^p\bigr)]=2i.
\ee
reproducing the results of the parallel branes results by the following idendifications:
\be
\left\{ \bea{cc}
\partial_{\theta}\alpha^p\mid_{\theta=0}=x^p \\ 
\alpha^p=p^p (momentum)\\
X^{(p+1)}\simeq \theta x^p \sigma.
\eea\right.
\ee
In this limit the low frequency spectrum turns into extra translational mode present
in the case of parallel branes.
  
  5. If we take more complicated relative orientations of the D-branes we will
 find further evidence for $\ \int J.A \ $ form of the interaction. 

{\it IV.P and $P'$ angled D-branes in type-II theories}: 

In this case we follow the method we have used in the previous cases. Since we don't
have $RR$ interactions, we expect the amplitude to be invariant under $\theta\rightarrow  1-\theta$.
In this case we have (31) but ${\bf R}$ and ${\bf NS}$ are replaced by:

\be
{\bf R}=(2\pi)^{3-\Delta/2} \biggl(\frac{\Theta_2(0\mid it)}{\Theta'_1(0\mid it)}\biggr)^{3-\Delta/2} \;\;\ 
\biggl(\frac{\Theta_3(0\mid it)}{\Theta_4(0\mid it)}\biggr)^{\Delta/2} \;\;\
\frac{i\Theta_2(it\theta \mid it)}{\Theta_1(it\theta \mid it)}
\ee
\be
{\bf NS}=(2\pi)^{3-\Delta/2} \biggl(\frac{\Theta_3(0\mid it)}{\Theta'_1(0\mid it)}\biggr)^{3-\Delta/2} \;\;\ 
\biggl(\frac{\Theta_2(0\mid it)}{\Theta_4(0\mid it)}\biggr)^{\Delta/2} \;\;\
\frac{i\Theta_3(it\theta \mid it)}{\Theta_1(it\theta \mid it)}
\ee
Then the massless contribution to amplitude reads as:
\be
 A(\theta)=-2^{\Delta}V_{p'}{F(\theta) \over \sin(\pi\theta)}(4\pi^2\alpha')^{3-(p+p')/2} \ G_{8-p}(Y^2).
\ee
\be
F(\theta)=3-\Delta+  \cos 2\pi\theta
\ee
This amplitude is similar to the gravitational interaction of D-brane in the bosonic case 
given in the appendix. When $P$ and $P'$ are different two branes can not detect
their $RR$ charge. Both $\theta=0$ and $\theta=1$ are stable points and the  
frequency of small oscillations around equilibrium points is proportional to 
${\sqrt{T_pT_{p'}}}=(4\pi^2\alpha')^{3-(p+p')/2}$.

It is worth noting the amplitude vanishes {\it at all orders of} $\ t\ $ for parallel branes when  $\Delta=4\ $ 
and for perpendicular branes when $\Delta=2$ .

{\it V.Moving branes}:                                                            

Moving D-brane result can be obtained by the following transformations from the 
angled D-branes :
\be 
 i\theta \rightarrow \epsilon \;\;\;\;\;\;\;\;\ \tanh \pi\epsilon=v.
\ee
where $v$ is the relative velocity of D-branes for both, the bosonic case or 
the type-II. This is exactly boosting one of the branes respect to the other one. 
If we do so the results are:

$P\ and\ P\ case:$
\be
\begin{array}{cc}
V_{RR}=8V_0 {1 \over \sqrt{1-v^2}}   \\
V_{NSNS}=-8V_0 \bigl( \frac{1-v^2/2}{1-v^2}\bigr)
\eea
\ee
$P\ and\ P'\ case:$ 
\be 
V_{NSNS}=-2V_0 \frac {4-\Delta-v^2(2-\Delta)}{1-v^2} \;\;\;\;\;\ V_{RR}=0. 
\ee
 They can be expected as "magnetic" forces induced by Lorentz transformations and
again can be justified by the field theoretic interaction term $\int J.A $.
 The $v^2$ dependence is due to two factors one from $J$ and the other from $A$.
In the limit of small velocities  we recover the results of [6],[4].

\vskip 1.5cm

As we have seen the relative orientation or velocity of the D-branes reveal more 
details about their structure and bringing them closer to the P-brane extended  
solutions of low energy SUGRA.

These correspondences make us believe that finally we are approching a totally
string theoretic language, a simple description of non-perturbative objects.

{\bf Acknowledgement:}

Authors would like to thank Porf. Ardalan for fruitful discussions.

\newpage
{\bf Appendix}:

Here we summerize the results for the bosonic  strings case in 26 dimensions:
$$
 A=2V_{p'+1} \int \frac{dt}{2t}(8\pi^2\alpha't)^{-(p'+1)/2}e^{-\frac{Y^2t}{2\pi^2\alpha'}}
 \bigl(q^{-2} \prod(1-q^{2n})^{-24}\bigr) $$
 \be
 \biggl(q^{1/8} \prod \frac{(1-q^{2n})}{(1-q^{2n-1})}\biggr)^{\Delta}
\ee
small $t$ limit:
\be
A= V_{p'+1}(24-2\Delta)(4\pi^2\alpha')^{11-{p+p' \over 2}}{\pi \over 2^{10}} 
G_{25-p}(Y^2)
\ee
Angled branes:
\be
\alpha'M^2=\frac{Y^2}{4\pi^2\alpha'}+N(\theta)-(1+\theta^2 /2) \;\;\;\;\
\ee
\be
N(\theta)=\sum_{n>0} \alpha_{-n}.\alpha_{n}
+\sum_{n>0}\alpha_{-n_+}^p.\alpha_{n_+}^p+ 
\sum_{n>0}\alpha_{-n_-}^p.\alpha_{n_-}^p
+{1 \over 2}( \alpha_{-\theta}^p \alpha_{\theta}^p+ 
\alpha_{\theta}^p \alpha_{-\theta}^p )
\ee
If we put these in $A$ it reads to be
\be
 A(\theta)=2V_{p} \int \frac{dt}{2t}(8\pi^2\alpha't)^{-p/2}e^{-\frac{Y^2t}{2\pi^2\alpha'}}
 \;\;\ \biggl({\Theta'_1(0\mid it) \over 2\pi}\biggr)^{-7} 
\frac{iq^{-\theta^2}}{\Theta_1(i\theta t \mid it)}  
\ee                                                                                                       

in small $t$ limit
\be
A(\theta)= V_{p}(4\pi^2\alpha')^{11-{p}}\ {\pi \over 2^{10}}\;\;\ F(\theta) 
{1 \over \sin\pi\theta }G_{24-p}(Y^2)
\ee
\be
F(\theta)=\bigl(22-4\sin^2\pi\theta  \bigr)
\ee
and small $t$ limit of the angled $P$ and $P'$ branes:
\be                                       
A(\theta)= V_{p'}(4\pi^2\alpha')^{11-{(p+p')/2}}\ {\pi \over 2^{10}}\;\;\ F(\theta) 
{1 \over \sin\pi\theta }G_{24-p}(Y^2)
\ee
\be
F(\theta)=\bigl(22-2\Delta-4\sin^2\pi\theta  \bigr) \;\;\;\;\ \Delta=p-p'.
\ee

\end{document}